\documentstyle[amstex,amssymb,preprint,aps,epsf,floats]{revtex}

\begin{document}

\tighten
\preprint{\vbox{
\hbox{DOE/ER/40762-215}
\hbox{UMD-PP-01-008}
}} \bigskip \bigskip

\title{Parity Violation in $\overrightarrow \gamma p$ Compton Scattering }
\author{Jiunn-Wei Chen, Thomas D. Cohen}
\address{Department of Physics, University of Maryland,\\
College Park, MD 20742-4111\\
{\tt jwchen@@physics.umd.edu, cohen@@physics.umd.edu}}
\author{Chung Wen Kao}
\address{Institut f\"{u}r Theoretische Physik, J. W. Goethe-Universit\"{a}t,\\
60054 Frankfurt am Main, Germany \\
{\tt kaochung@@th.physik.uni-frankfurt.de}}
\maketitle

\begin{abstract}
Polarized beam $\overrightarrow{\gamma }p$ Compton scattering provides a
theoretically clean way to extract the isovector parity violating
pion-nucleon coupling constant $h_{\pi NN}^{(1)}$. This channel is more
tractable experimentally than the recently proposed extraction of $h_{\pi
NN}^{(1)}$ from the Bedaque-Savage process --- polarized target $\gamma 
\overrightarrow{p}$ compton scattering. The leading parity violating effect
is calculated using Heavy Baryon Chiral Perturbation Theory $(HB\chi PT)$.
The size of the asymmetry is estimated to be $\sim 4\times 10^{-8}$ 
for 120 MeV photon energy.
\end{abstract}

\tighten
\bigskip \vskip3.0cm \leftline{} 





\vfill\eject
The isovector parity violating (PV) pion-nucleon coupling constant $h_{\pi
NN}^{(1)}$ is responsible for the longest range part of the $\Delta I=1$ PV $%
NN$ forces \cite{DDH,AH,KS}. It is expected to give dominant contributions
to low energy quantities such as nucleon\footnote{
The isoscalar nucleon anapole moment is dominated by $h_{\pi
NN}^{(1)}$, but not the isovector anapole moment which is more directly
relevant for the SAMPLE electron-proton~\cite{SAMPLEp} and 
electron-deuteron~\cite{SAMPLEd} PV experiments.} 
and nuclear anapole moment \cite
{ZE,HHM,MH,SR,vanKolck,ZPHM}, and PV neutron radiative capture $%
np\rightarrow d\gamma $. However, past attempts to extract $h_{\pi NN}^{(1)}$
are not satisfactory (see \cite{AH,HH,Oers} reviews). In many-body systems,
several PV effects are enhanced and have been detected. On the other hand,
the theoretical analysis is complicated. The extractions from $^{18}F$ \cite
{F18Page,F18Bini} and $^{133}Cs$ \cite{Cs123,Haxton,FM,WB} systems differ by
an order of magnitude with large uncertainties, while the measurement in the 
$^{205}Tl$ system gives a null result \cite{Tl205}. In fewer body systems,
the theory is more under control but the PV effect is smaller such that
previous measurements could not reach the required precision \cite
{npcirgam,PVnpdgam,PVgamd,Markoff}. However there are several high precision
new measurements under preparation or execution including $\overrightarrow{n}%
p\rightarrow d\gamma $ at LANSCE \cite{LANSCE}, $\gamma d\rightarrow np$ at
JLab \cite{JLab}, and the rotation of polarized neutrons in helium at NIST.
It is expected that these experiments will put tight constraints on the
value of $h_{\pi NN}^{(1)}$.

In the single nucleon system, a new PV observable was recently suggested by
Bedaque and Savage\cite{BS}. They found that the polarized target $\gamma 
\overrightarrow{p}\rightarrow \gamma p$ Compton scattering asymmetry
measurement (calculated to be $\sim 5\times 10^{-8}$ for 100 MeV photon
energy) would determine $h_{\pi NN}^{(1)}$ with an estimated 15\%
uncertainty. To control systematic errors, the difference in cross section
for the proton spin polarized parallel and antiparallel to the direction of
the incident photon must be measured during a short period of time with only
the target polarization direction changed. To achieve this, the rapid
flipping of the target polarization should be employed. This is unpractical
with the currently available experimental techniques and polarized proton
targets, thus the Bedaque-Savage (BS) process is not favored experimentally.
On the other hand, rapid flipping of beam helicity is a standard technique
already employed in many parity violating experiments. Thus the polarized
beam experiment, $\overrightarrow{\gamma }p\rightarrow \gamma p$, is
experimentally more tractable \cite{RS}. Given the great interest in the
determination of $h_{\pi NN}^{(1)}$, we investigate this $\overrightarrow{%
\gamma }p$ parity violating process using Heavy Baryon Chiral Perturbation
Theory ($HB\chi PT$) \cite{HBChPT,BKM}.

We start with reviewing the symmetry constraints on the Compton scattering
process 
\begin{equation}
\gamma \left( k,\epsilon \right) +p\rightarrow \gamma \left( k^{\prime
},\epsilon ^{\prime }\right) +p\qquad ,
\end{equation}
where $\left( k,\epsilon \right) $ and $\left( k^{\prime },\epsilon ^{\prime
}\right) $ are the initial and final photon momenta and polarization
vectors. It has been known for a long time that there are ten time reversal
invariant structure functions in the transition amplitude for this process.
Six of them conserve parity while four of them violate parity. This can be
seen easily by the following exercise on helicity amplitude counting. Using $%
\left| \lambda _{1},\lambda _{2}\right\rangle $ to denote a state with
photon and proton helicity $\lambda _{1}$ and $\lambda _{2}$ in the center
of mass frame, sixteen helicity amplitudes $\left\langle \lambda
_{1}^{\prime },\lambda _{2}^{\prime }|\lambda _{1},\lambda _{2}\right\rangle 
$ can be constructed for proton Compton scattering. These amplitudes
transform under time reversal as 
\begin{equation}
\left\langle \lambda _{1}^{\prime },\lambda _{2}^{\prime }|\lambda
_{1},\lambda _{2}\right\rangle \stackrel{T}{\rightarrow }\left\langle
\lambda _{1},\lambda _{2}|\lambda _{1}^{\prime },\lambda _{2}^{\prime
}\right\rangle \qquad ,
\end{equation}
which is equivalent to taking a transpose transformation of the 4$\times 4$
matrix. Thus ten time reversal invariant amplitudes can be constructed
through the linear combination 
\begin{equation}
{\cal M}_{T}(\lambda _{1},\lambda _{2},\lambda _{1}^{\prime },\lambda
_{2}^{\prime })=\left\langle \lambda _{1}^{\prime },\lambda _{2}^{\prime
}|\lambda _{1},\lambda _{2}\right\rangle +\left\langle \lambda _{1},\lambda
_{2}|\lambda _{1}^{\prime },\lambda _{2}^{\prime }\right\rangle \qquad .
\label{ampT}
\end{equation}
\ One can further separate these into PC and PV amplitudes. Two of the
amplitudes, ${\cal M}_{T}(+1,+1/2,-1,-1/2)$ and ${\cal M}%
_{T}(+1,-1/2,-1,+1/2),$ have an additional symmetry. They are invariant
under parity transformation, 
\begin{equation}
\left\langle \lambda _{1}^{\prime },\lambda _{2}^{\prime }|\lambda
_{1},\lambda _{2}\right\rangle \stackrel{P}{\rightarrow }\left\langle
-\lambda _{1}^{\prime },-\lambda _{2}^{\prime }|-\lambda _{1},-\lambda
_{2}\right\rangle \qquad ,
\end{equation}
after having been made to conserve time reversal invariance. The other eight
amplitudes can be grouped into four PC and four PV amplitudes using similar
linear combinations to that of eq.(\ref{ampT}). This demonstrates the
well-known result that there are six independent PC and four independent PV
structure functions satisfying time-reversal invariance in a proton Compton
scattering process.

In the center of mass frame, the six PC structure functions can be chosen as 
\begin{eqnarray}
T^{pc} &=&\overline{N}\left[ {\cal A}_{1}\,{\bf \epsilon }\cdot {\bf %
\epsilon }^{\prime \ast }+{\cal A}_{2}\widehat{{\bf k}}\cdot {\bf \epsilon }%
^{\prime \ast }\,\widehat{{\bf k}^{\prime }}\cdot {\bf \epsilon }+i{\cal A}%
_{3}{\bf \sigma }\cdot ({\bf \epsilon }^{\prime \ast }\times \,{\bf \epsilon 
})+i{\cal A}_{4}{\bf \sigma }\cdot (\,\widehat{{\bf k}^{\prime }}\times 
\widehat{{\bf k}}){\bf \epsilon }\cdot {\bf \epsilon }^{\prime \ast }\right. 
\nonumber \\
&&+i{\cal A}_{5}{\bf \sigma }\cdot \left[ ({\bf \epsilon }^{\prime \ast
}\times \widehat{{\bf k}}){\bf \epsilon }\cdot \widehat{{\bf k}^{\prime }}-(%
{\bf \epsilon }\times \widehat{{\bf k}^{\prime }}){\bf \epsilon }^{\prime
\ast }\cdot \widehat{{\bf k}}\right]   \nonumber \\
&&\left. +i{\cal A}_{6}{\bf \sigma }\cdot \left[ ({\bf \epsilon }^{\prime
\ast }\times \widehat{{\bf k}^{\prime }}){\bf \epsilon }\cdot \widehat{{\bf k%
}^{\prime }}-({\bf \epsilon }\times \widehat{{\bf k}}){\bf \epsilon }%
^{\prime \ast }\cdot \widehat{{\bf k}}\right] \right] N\qquad ,
\end{eqnarray}
where $N$ is the proton spinor, ${\bf \sigma }$ is the Pauli matrix acting
on the nucleon spin index, $\widehat{{\bf k}}$ and $\widehat{{\bf k}^{\prime
}}$ are the unit vectors in the ${\bf k}$ and ${\bf k}^{\prime }$
directions, and the Coulomb gauge (${\bf \epsilon }_{0}={\bf \epsilon }%
_{0}^{\prime }=0)$ is used. The PV structure functions can be chosen as 
\begin{eqnarray}
T^{pv} &=&\overline{N}\left[ {\cal F}_{1}\,{\bf \sigma }\cdot (\widehat{{\bf %
k}}+\widehat{{\bf k}^{\prime }})\,{\bf \epsilon }\cdot {\bf \epsilon }%
^{\prime \ast }-{\cal F}_{2}\left( {\bf \sigma }\cdot {\bf \epsilon }%
^{\prime \ast }\,\widehat{{\bf k}^{\prime }}\cdot {\bf \epsilon }+{\bf %
\sigma }\cdot {\bf \epsilon \,}\widehat{{\bf k}}\cdot {\bf \epsilon }%
^{\prime \ast }\right) \right.   \nonumber \\
&&\left. -{\cal F}_{3}\widehat{{\bf k}}\cdot {\bf \epsilon }^{\prime \ast }\,%
\widehat{{\bf k}^{\prime }}\cdot {\bf \epsilon \,\sigma }\cdot (\widehat{%
{\bf k}}+\widehat{{\bf k}^{\prime }})-i{\cal F}_{4}{\bf \epsilon }\times 
{\bf \epsilon }^{\prime \ast }\cdot (\widehat{{\bf k}}+\widehat{{\bf k}%
^{\prime }})\right] N\qquad .
\end{eqnarray}
The ${\cal F}_{1\text{-}3}$ structures were first given in ref.\cite{BS}.
The interference between ${\cal A}_{1,2}$ and ${\cal F}_{1\text{-}3}$
contributes to the BS process. For the polarized beam process $%
\overrightarrow{\gamma }p\rightarrow \gamma p$ considered here, the
contributions are from the interference between ${\cal A}_{1,2}$ and ${\cal F%
}_{4}$ and between ${\cal A}_{3\text{-}6}$ and ${\cal F}_{1\text{-}3}$.

Since we are interested in the low energy behavior of the proton Compton
scattering process, chiral perturbation theory provides a natural framework
with which to work. As a low energy effective field theory of QCD, chiral
perturbation theory captures the symmetries of QCD and describes low energy
observables by derivative and Chiral expansions. The $SU(2)_{L}\times U(1)$
symmetry structure of electroweak interactions can also be incorporated with
the weak boson exchange described by contact interactions while keeping the
photon as dynamical degrees of freedom in the Chiral Lagrangian.

\begin{figure}[t]
\centerline{\epsfxsize=4.0in \epsfbox{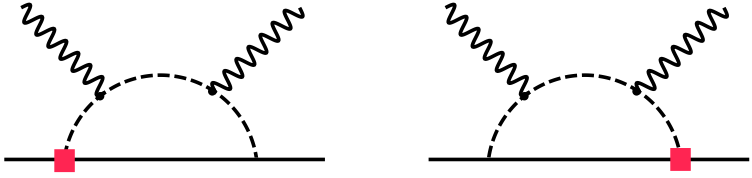}}
\noindent
\caption{{\it The leading order contribution to parity violating sturcture
functions ${\cal F}_{1}$-${\cal F}_{3}$ in $\protect\gamma p$ Compton
scattering. The solid square is the weak operator with coefficient $h_{%
\protect\pi NN}^{(1)}$. Wavy lines are photons, solid lines are nucleons,
and dashed lines are pions. The crossed graphs are not shown. Graphs with
photons from the strong vertex, or insertion of the two-photon-pion vertex
vanish in the $v.A=0$ gauge, and thus are not shown here. }}
\label{figf123}
\end{figure}

\begin{figure}[t]
\centerline{\epsfxsize=4.0in \epsfbox{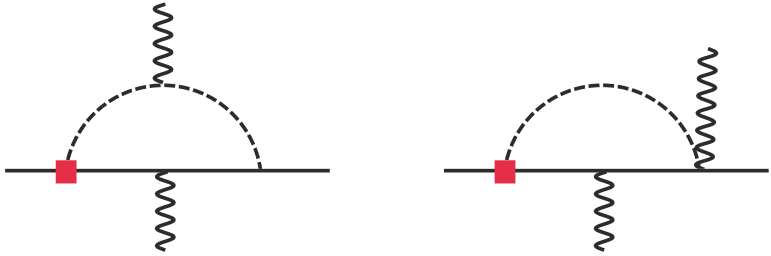}}
\noindent
\caption{{\it The first non-vanishing order (NLO) contribution to parity
violating sturcture functions ${\cal F}_{4}$ in $\protect\gamma p$ Compton
scattering. The features of the graphs are as defined in fig.\ref{figf123}.
The photon nucleon couplings are magntic couplings. } }
\label{figf4}
\end{figure}

$HB\chi PT$ has been applied to the calculation of several Compton
scattering observables. The PC structure functions have been calculated up
to next-to-leading order (NLO), ${\cal O}(e^{2}p)$ \cite{BKM}, and are
listed in the Appendix. The proton Thompson term is the only contribution at
leading order (LO) and contributes to ${\cal A}_{1}$ only. The NLO
contributions come from the tree level diagrams, pion-nucleon loop diagrams
and the Wess-Zumino term. They contribute to all the PC structure functions $%
{\cal A}_{1\text{-}6}$. These structure functions can be determined
experimentally (see the Appendix for more details) thus the uncertainty from
the PC part can be eliminated completely. Here we use the $HB\chi PT$ \
result only for the sake of estimation of the asymmetry. For the PV
structure functions, the LO (${\cal O}(G_{F}e^{2})$ with $G_{F}$ the Fermi
coupling constant) contributions have been calculated in ref.\cite{BS}. We
also list them in the Appendix. They arise from the pion loop diagrams shown
in fig.\ref{figf123} and contribute to ${\cal F}_{1}$-${\cal F}_{3}.$ ${\cal %
F}_{4}$ is an additional quantity we will need to compute. Its contribution
comes from the pion loop diagrams shown in fig.\ref{figf4} and the effect
starts at NLO. Now we give some details of computing the ${\cal F}_{4}^{NLO}$%
.

%

The PC part of the relevant Lagrangian is

\begin{eqnarray}
{\cal L}^{PC} &=&\frac{1}{2}D_{\mu }\pi _{i}D^{\mu }\pi _{i}-\frac{m_{\pi
}^{2}}{2}\pi _{i}^{2}+i\overline{N}v_{\mu }D^{\mu }N-\frac{g_{A}}{F_{\pi }}%
\overline{N}S_{\mu }(D^{\mu }{\cal \pi }_{i})\tau _{i}N  \nonumber \\
&&+\ \frac{1}{2M_{N}}\overline{N}\left[ \left( v\cdot D\right) ^{2}-D^{2}%
\right] N-\frac{ie}{M_{N}}\overline{N}\left[ S^{\mu },S^{\nu }\right] \left[
\mu _{0}+\mu _{1}\tau _{3}\right] NF_{\mu \nu }+\cdots 
\end{eqnarray}
where the pion decay constant $F_{\pi }=93$ MeV, pion-nucleon coupling
constant $g_{A}=1.26$, $N$ is the isospin doublet of the nucleon fields with
velocity $v$, $D$ is the covariant derivative with gauge coupling on the
proton as $D_{\mu }N=(\partial -ieA)_{\mu }N,$ and $S$ is the covariant
nucleon polarization vector. In the proton rest frame, $v^{\mu }=(1,0,0,0)$, 
$S^{\mu }=(0,{\bf \sigma }/2)$. $\mu _{0}=(\mu _{p}+\mu _{n})/2$ and $\mu
_{1}=(\mu _{p}-\mu _{n})/2$ are the isoscalar and isovector magnetic moments
in nuclear magnetons, with $\mu _{p}=2.79$ and $\mu _{n}=-1.91.$ The
ellipses denote terms with more pion fields and insertions of higher powers
of derivative and pion mass. Massive hadronic excitations such as kaons and
deltas are ``integrated out''. Their effects are encoded in the higher
dimensional operators.

The non-leptonic PV part of the relevant Lagrangian is 
\begin{equation}
{\cal L}^{PV}=\frac{h_{\pi NN}^{(1)}}{\sqrt{2}}\varepsilon ^{3ij}\overline{N}%
\pi _{i}\tau _{j}N+\cdots =-ih_{\pi NN}^{(1)}\pi ^{+}p^{\dagger
}n+h.c.+\cdots   \label{Lw}
\end{equation}
where the ellipses denote terms with more pion fields and derivatives. This
Lagrangian was first given in ref.\cite{KS} with a different phase convention
for the pion field. We adopt the same convention as refs.\cite
{vanKolck,ZPHM}. $h_{\pi NN}^{(1)}$ was estimated by matching onto four
quark Fermi theory and was found to be dominated by $s$ quark contributions, 
\begin{equation}
\left| h_{\pi NN}^{(1)}\right| \sim \frac{G_{F}F_{\pi }\Lambda _{\chi }}{%
\sqrt{2}}\sim 5\times 10^{-7}\qquad ,  \label{hpinn}
\end{equation}
where $\Lambda _{\chi }\sim 1$ GeV is the chiral perturbation scale. This
estimation is consistent with the ``best value'' obtained in ref.\cite{DDH}
and close to one result\cite{HHK} from QCD sum rules. 
A recent calculation in the SU(3) Skyrme model yields $h_{\pi
NN}^{(1)}\sim 0.8$-$1.3$ $\times 10^{-7}$\cite{MW}. The 
radiative correction on the $h_{\pi NN}^{(1)}$ vertex is discussed in 
ref.\cite{ZPHM2}.

The $h_{\pi NN}^{(1)}$ term is the only term in the ${\cal L}^{PV}$ with a
non-derivative pion-nucleon coupling; it is expected to dominate over the
other contributions to ${\cal F}_{4}.$ The pion loop diagrams in fig.\ref
{figf4} gives 
\begin{equation}
{\cal F}_{4}^{NLO}=-%
{\displaystyle{e^{2}g_{A}h_{\pi NN}^{(1)}\mu _{n} \over 8\sqrt{2}m_{N}F_{\pi }\pi ^{2}}}%
\left[ \omega -\frac{m_{\pi }^{2}}{\omega }\left( \sin ^{-1}\frac{\omega }{%
m_{\pi }}\right) ^{2}\right] \qquad
\end{equation}
with a magnetic photon-nucleon coupling. $\omega$ is the photon energy in the 
center-of-mass frame.

In this calculation delta contributions are encoded in the higher-order
operators and will contribute through higher-order diagrams. If one
considers the delta-nucleon mass difference $\Delta \sim 300$ MeV as a light
scale as $m_{\pi }$ and \ $\omega $ are, then one needs to sum factors of $%
m_{\pi }/\Delta $ and $\ \omega /\Delta $ to all orders. This can be done by
including delta as a dynamical degree of freedom \cite{HBChPT,Hemmert}. In
this expansion, the delta diagrams will contribute to ${\cal F}_{1-4}^{NLO}$
through $\pi \Delta $ loop diagrams and tree diagrams\ with unknown $\gamma
N\Delta $ couplings. In the expansion with which we work, the $\Delta $ is
considered as a large or heavy scale, so factors of $m_{\pi }/\Delta $ and $%
\omega /\Delta $\ are treated perturbatively. Thus below pion production
threshold ($\omega <m_{\pi }),$ the delta would contribute a factor of $%
(m_{\pi }^{2}/\Delta ^{2},\omega ^{2}/\Delta ^{2})\sim 25\%$ correction to $%
{\cal F}_{4}^{NLO}$. This is the dominant source of the uncertainty.

The PV asymmetry can be defined by the difference in the cross section (in
the center-of-mass frame) for photon helicity $\lambda _{\gamma }=+1$ and $-1
$ normalized to the sum 
\begin{equation}
A_{\gamma \gamma }\left( \omega ,\theta \right) \equiv \frac{%
{\displaystyle{d\sigma  \over d\Omega }}%
(\lambda _{\gamma }=+1)-%
{\displaystyle{d\sigma  \over d\Omega }}%
(\lambda _{\gamma }=-1)}{%
{\displaystyle{d\sigma  \over d\Omega }}%
(\lambda _{\gamma }=+1)+%
{\displaystyle{d\sigma  \over d\Omega }}%
(\lambda _{\gamma }=-1)}\qquad .  \label{eq:asym}
\end{equation}
Again, the PC structure functions ${\cal A}_{1\text{-}6}$ can be extracted
from experiments to further reduce theoretical input and eliminate the
uncertainties from the PC part. Here we plug in the PC $HB\chi PT$ \ result
in order to get an estimation of the size of the asymmetry. In this
treatment, the helicity asymmetry contribution starts at NLO, 
\begin{eqnarray}
A_{\gamma \gamma }\left( \omega ,\theta \right)  &=&\frac{2\sin ^{2}\theta }{%
\left| {\cal A}_{1}^{LO}\right| ^{2}(1+\cos ^{2}\theta )}%
\mathop{\rm Re}%
\left\{ {\cal A}_{1}^{LO}{\cal F}_{4}^{NLO^{\ast }}+{\cal A}_{3}^{NLO}\left[ 
{\cal F}_{1}^{LO^{\ast }}-2{\cal F}_{2}^{LO^{\ast }}-{\cal F}_{3}^{LO^{\ast
}}\left( 1+\cos \theta \right) \right] \right.   \nonumber \\
&&+{\cal A}_{5}^{NLO}\left[ {\cal F}_{1}^{LO^{\ast }}\left( 1+\cos \theta
\right) +{\cal F}_{2}^{LO^{\ast }}\left( 1-3\cos \theta \right) +{\cal F}%
_{3}^{LO^{\ast }}\left( 1-\cos ^{2}\theta \right) \right]   \nonumber \\
&&-{\cal A}_{6}^{NLO}\left[ {\cal F}_{1}^{LO^{\ast }}\left( 1+\cos \theta
\right) +{\cal F}_{2}^{LO^{\ast }}\left( 3-\cos \theta \right) +{\cal F}%
_{3}^{LO^{\ast }}\left( 1-\cos ^{2}\theta \right) \right]   \nonumber \\
&&\left. -{\cal A}_{4}^{NLO}{\cal F}_{2}^{LO^{\ast }}\left( 1+\cos \theta
\right) \right\}   \nonumber \\
&&-\frac{4(1+\cos \theta )}{\left| {\cal A}_{1}^{LO}\right| ^{2}(1+\cos
^{2}\theta )}%
\mathop{\rm Re}%
\left\{ \ {\cal A}_{3}^{NLO}{\cal F}_{1}^{LO^{\ast }}+\ {\cal A}_{1}^{LO}%
{\cal F}_{4}^{NLO^{\ast }}\right\} \qquad .  \label{asym}
\end{eqnarray}
We have used the LO result for the PC cross section, so 
\begin{equation}
{\displaystyle{1 \over 2}}%
\left[ 
{\displaystyle{d\sigma  \over d\Omega }}%
(\lambda _{\gamma }=+1)+%
{\displaystyle{d\sigma  \over d\Omega }}%
(\lambda _{\gamma }=-1)\right] =%
{\displaystyle{1 \over 2\left( 4\pi \right) ^{2}}}%
\left| {\cal A}_{1}^{LO}\right| ^{2}(1+\cos ^{2}\theta )=%
{\displaystyle{\alpha ^{2} \over 2M_{N}^{2}}}%
(1+\cos ^{2}\theta )\qquad .
\end{equation}

It is instructive to study the low energy limit $\left( \omega \ll m_{\pi
}\right) $\ of the asymmetry. Keeping the first term in the $\omega /m_{\pi }
$ expansion, 
\begin{eqnarray}
{\cal A}_{1}^{LO} &=&-\frac{e^{2}}{M_{N}}\qquad ,\qquad {\cal A}%
_{3}^{NLO}\sim \left[ 1+2\kappa _{p}-\left( 1+\kappa _{p}\right) ^{2}\cos
\theta \right] \frac{e^{2}\omega }{2M_{N}^{2}}\qquad ,  \nonumber \\
{\cal A}_{4}^{NLO} &=&-{\cal A}_{5}^{NLO}\sim -\frac{\left( 1+\kappa
_{p}\right) ^{2}e^{2}\omega }{2M_{N}^{2}}\qquad ,\qquad {\cal A}%
_{6}^{NLO}\sim -\frac{\left( 1+\kappa _{p}\right) e^{2}\omega }{2M_{N}^{2}}%
\qquad ,  \nonumber \\
{\cal F}_{1}^{LO^{\ast }} &=&{\cal F}_{2}^{LO^{\ast }}\sim -\frac{%
e^{2}g_{A}h_{\pi NN}^{(1)}\omega ^{2}}{24\sqrt{2}\pi ^{2}F_{\pi }m_{\pi }^{2}%
}\qquad ,\qquad {\cal F}_{4}^{NLO^{\ast }}\sim +\frac{e^{2}g_{A}h_{\pi
NN}^{(1)}\mu _{n}\omega ^{3}}{24\sqrt{2}\pi ^{2}F_{\pi }M_{N}m_{\pi }^{2}}%
\qquad ,
\end{eqnarray}
with ${\cal F}_{3}^{LO^{\ast }}=O(\omega ^{4}/m_{\pi }^{4})$ and $\kappa
_{p}\equiv \mu _{p}-1$. The inverse power of $m_{\pi }$ dependence in the $%
{\cal F}$s explains why there are no intrinsic unknown
two-photon-two-nucleon counterterms at this order. In this low energy limit,
the asymmetry has a simple form 
\begin{eqnarray}
A_{\gamma \gamma }\left( \omega \ll m_{\pi },\theta \right)  &=&-\frac{%
g_{A}h_{\pi NN}^{(1)}\left[ \left( 2\mu _{n}+\left( \mu _{p}+1\right)
^{2}\right) \sin ^{2}\theta -2\left( 1+\cos \theta \right) \left( 2\mu
_{n}-\left( \mu _{p}-1\right) ^{2}\right) \right] \omega ^{3}}{24\sqrt{2}\pi
^{2}F_{\pi }m_{\pi }^{2}(1+\cos ^{2}\theta )} \\
&&\cdot \left( 1+{\cal O}\left( \frac{\omega ^{2}}{m_{\pi }^{2}},\frac{%
m_{\pi }^{2}}{\Delta ^{2}}\right) \right) \qquad .
\end{eqnarray}
The vanishing of $A_{\gamma \gamma }$ at backward angle ($\theta =\pi $) is
a consequence of time-reversal invariance and is a general property for all
values of the photon energy. For back scattering, the change of photon spin
direction corresponds to a $\Delta J=2$ operation and hence, is a forbidden
transition for the proton matrix element. Thus both photon and proton spin
directions do not change but the helicities change signs. Using eq.(\ref
{ampT}) for the time-reversal invariant amplitude and $\left( \lambda
_{1}^{\prime },\lambda _{2}^{\prime }\right) =$ $\left( -\lambda
_{1},-\lambda _{2}\right) $, this amplitude conserves parity and does not
contribute to the PV asymmetry.

For a numerical estimation of the magnitude of the asymmetry, we consider $%
\theta =\pi /2$, where 
\begin{equation}
A_{\gamma \gamma }\left( \omega \ll m_{\pi },\frac{\pi }{2}\right) \sim
-8.8\times 10^{-9}\left( \frac{h_{\pi NN}^{(1)}}{5\times 10^{-7}}\right)
\left( \frac{\omega }{70MeV}\right) ^{3}
\end{equation}
with $\sim 25\%$ uncertainty.

\begin{figure}[t]
\centerline{\epsfxsize=4.0in \epsfbox{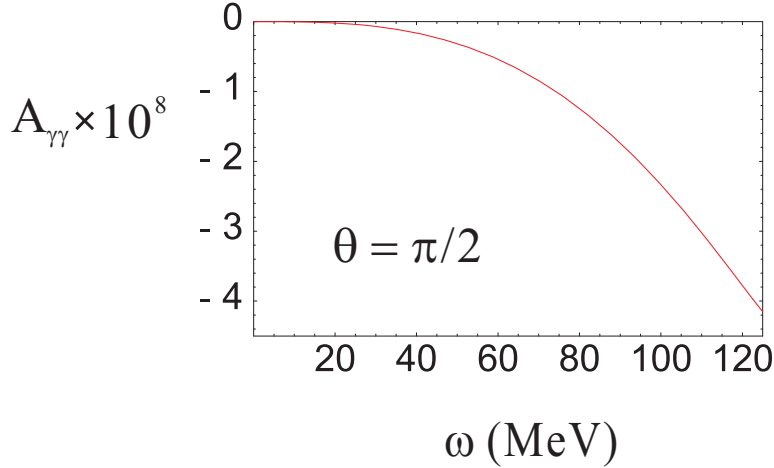}}
\noindent
\vskip0.3in
\caption{{\it The estimated photon helicity asymmetry $A_{\protect\gamma 
\protect\gamma }$ defined in eq.(\ref{eq:asym}) shown as a function of photon
energy with the photon reflection angle $\protect\theta =\protect\pi /2$ in
the center-of-mass-frame. The naively estimated value of $h_{\protect\pi
NN}^{(1)}=5\times 10^{-7}$ is taken as input, and a HB$\chi$PT estimation used
for the parity conserving amplitude. }}
\label{fig:asym}
\end{figure}

In fig.\ref{fig:asym}, we show the photon energy dependence of the estimated
asymmetry at $\theta =\pi /2.$ Assuming the naive size for $h_{\pi NN}^{(1)}$
estimated in eq.(\ref{hpinn}), the asymmetry is $A_{\gamma \gamma }\left( 120%
\text{ MeV},\pi /2\right) \sim -3.8\times 10^{-8}$, with the higher-order
uncertainty $\sim $($m_{\pi }^{2}/\Delta ^{2},\omega ^{2}/\Delta ^{2})\sim
25\%.$ Note that very near the pion production threshold, resummation of
terms with powers of 
\begin{equation}
\frac{m_{\pi }^{2}}{2m_{N}\left( \omega -m_{\pi }\right) }  \label{resum}
\end{equation}
is required in order to shift the pion production threshold from $m_{\pi }$
to $m_{\pi }+%
{\displaystyle{m_{\pi }^{2} \over 2m_{N}}}%
$ (in the laboratory frame) to recover the recoil effect. The resummation
procedure is well known \cite{BL}. Without this resummation, we should
restrict ourselves to $\omega <130$ MeV such that the factor in (\ref{resum}%
) is sufficiently less than 1. To probe how far away from the threshold our
calculation is still under control, we compare the results of expanding the $%
A_{\gamma \gamma }$ to NLO to that of expanding the total amplitude to NLO,
then square it to get the $A_{\gamma \gamma }$ (the ${\cal A}_{2}$ ${\cal F}%
_{4}$ interference term is included in the latter case). If the difference
between these two quantities is consistent with the estimated higher order
contribution, then the result without extra resummation will be valid. We
find the difference increases with $\omega $ and reaches 14\% at $\omega =120
$ MeV, so our expansion is presumably still useful up to 120 MeV.

\begin{figure}[t]
\centerline{\epsfxsize=4.0in \epsfbox{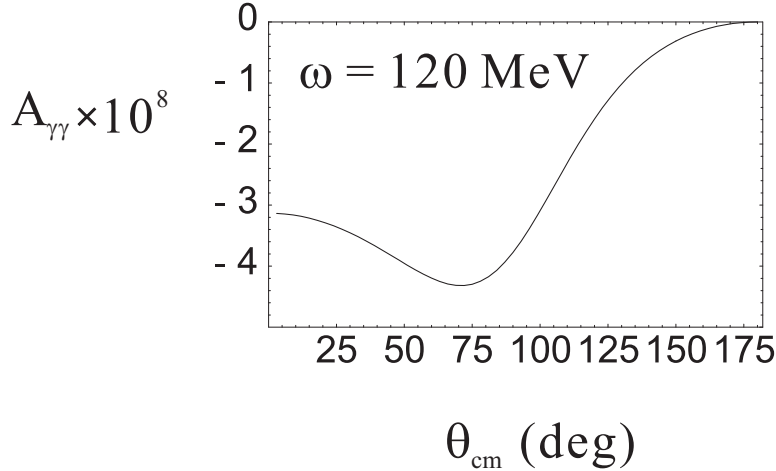}}
\noindent
\vskip0.3in
\caption{{\it The angular distribution of the estimated photon helicity
asymmetry $A_{\protect\gamma \protect\gamma}$ defined in eq.(\ref{eq:asym}) 
and calculated in HB$\chi$PT in
the center-of-mass-frame for 100 MeV photon energy. The naively estimated
value of $h^{(1)}_{\protect\pi N N}=5 \times 10^{-7}$ is taken as input. }}
\label{fig:angdis}
\end{figure}

In fig.\ref{fig:angdis}, we show the angular distribution at $\omega =120$
MeV for the same estimated value of $h_{\pi NN}^{(1)}$. The maximum
asymmetry is near $\theta =\pi /2$ but slightly biased towards the forward
direction. The asymmetry vanishes at $\theta =\pi $ as required by time
reversal invariance.

In conclusion, parity violating $\overrightarrow{\gamma }p\rightarrow \gamma
p$ Compton scattering provides a theoretically clean way to extract $h_{\pi
NN}^{(1)}$. The dominating source of the PV effect comes from the PV pion
loop contributions. The magnitude of the helicity asymmetry is estimated to
be $\sim 4\times 10^{-8}$ at 120 MeV photon energy with $\sim $25\%
uncertainty for a natural size $h_{\pi NN}^{(1)}$, under the framework of HB$%
\chi $PT. Thus we have found a model independent way to\ constrain $h_{\pi
NN}^{(1)}$ with $\sim $25\% uncertainty. We note that the $\sim $25\%
uncertainty is dominantly due to the delta and can in principle be reduced
by inclusion of the delta as an explicit degree of freedom. Unfortunately
this would require additional experiments to measure the PV $\gamma N\Delta $
coupling. However, even with $\sim $25\% uncertainties, $\overrightarrow{%
\gamma }p$ Compton scattering will greatly improve our understanding of $%
h_{\pi NN}^{(1)}$. We are optimistic that this experiment is feasible for
current experimental techniques and facilities.

\vskip1in \centerline{\bf ACKNOWLEDGMENTS} \bigskip

J.W.C. and T.D.C. thank the Institute for Nuclear Theory at the University
of Washington for its hospitality and the Department of Energy for partial
support during the completion of this work. J.W.C. thanks P. Bedaque, T.
Hemmert, X. Ji, U. Meissner, M. Ramsey-Musolf, M. Savage, R. Springer, and
R. Suleiman for useful discussions. J.W.C. and T.D.C. are supported in part
by the U.S. Dept. of Energy under grant No. DE-FG02-93ER-40762. C.W.K is
supported by the Alexander von Humboldt Foundation.

\vfill
\vskip1in \centerline{\bf Appendix} \bigskip

The following PC structure functions computed to ${\cal O}(e^{2}p)$ are
taken from eqs. (4.28a-g) of ref.\cite{BKM} (with a typo in eq.(4.28b)
corrected). 
\begin{eqnarray}
{\cal A}_{1}^{LO} &=&-\frac{e^{2}}{M_{N}}  \nonumber \\
{\cal A}_{1}^{NLO} &=&{\frac{g_{A}^{2}e^{2}}{8\pi F_{\pi }^{2}}}\left\{
m_{\pi }-\sqrt{m_{\pi }^{2}-\omega ^{2}}+\frac{2M_{\pi }^{2}-t}{\sqrt{-t}}%
\left[ \frac{1}{2}\arctan {\frac{\sqrt{-t}}{2m_{\pi }}}\right. \right.  
\nonumber \\
&&\left. \left. -\int_{0}^{1}dz\arctan {\frac{(1-z)\sqrt{-t}}{2\sqrt{M_{\pi
}^{2}-\omega ^{2}z^{2}}}}\right] \right\} 
\end{eqnarray}

\begin{equation}
{\cal A}_{2}^{NLO}={\frac{e^{2}\omega }{M_{N}^{2}}}+{\frac{%
e^{2}g_{A}^{2}\omega ^{2}}{8\pi F_{\pi }^{2}}}{\frac{t-2m_{\pi }^{2}}{%
(-t)^{3/2}}}\int_{0}^{1}dz\left[ \arctan {\frac{(1-z)\sqrt{-t}}{2\sqrt{%
m_{\pi }^{2}-\omega ^{2}z^{2}}}}-{\frac{2(1-z)\sqrt{t(\omega
^{2}z^{2}-m_{\pi }^{2})}}{4m_{\pi }^{2}-4\omega ^{2}z^{2}-t(1-z)^{2}}}\right]
\end{equation}
\begin{eqnarray}
{\cal A}_{3}^{NLO} &=&\frac{e^{2}\omega }{2M_{N}^{2}}\left[ 1+2\kappa
_{p}-(1+\kappa _{p})^{2}\cos \theta \right] +\frac{e^{2}g_{A}t\omega }{8\pi
^{2}F_{\pi }^{2}(m_{\pi }^{2}-t)}+{\frac{e^{2}g_{A}^{2}}{8\pi ^{2}F_{\pi
}^{2}}}\left[ \frac{M_{\pi }^{2}}{\omega }\arcsin ^{2}\frac{\omega }{m_{\pi }%
}-\omega \right]  \nonumber \\
&&+{\frac{e^{2}g_{A}^{2}}{4\pi ^{2}F_{\pi }^{2}}\omega ^{4}}\sin ^{2}\theta
\int_{0}^{1}dx\int_{0}^{1}dz\frac{x(1-x)z(1-z)^{3}}{W^{3}}\left[ \arcsin 
\frac{\omega z}{R}+\frac{\omega zW}{R^{2}}\right]
\end{eqnarray}

\begin{equation}
{\cal A}_{4}^{NLO}=-{\frac{e^{2}(1+\kappa _{p})^{2}\omega }{2M_{N}^{2}}}+{%
\frac{e^{2}g_{A}^{2}}{4\pi ^{2}F_{\pi }^{2}}}\int_{0}^{1}dx\int_{0}^{1}dz{%
\frac{z(1-z)}{W}}\arcsin {\frac{\omega z}{R}}
\end{equation}
\begin{eqnarray}
{\cal A}_{5}^{NLO} &=&\frac{e^{2}\omega }{2M_{N}^{2}}(1+\kappa _{p})^{2}-%
\frac{e^{2}g_{A}\omega ^{3}}{8\pi ^{2}F_{\pi }^{2}(m_{\pi }^{2}-t)}+{\frac{%
e^{2}g_{A}^{2}\omega ^{2}}{8\pi ^{2}F_{\pi }^{2}}}\int_{0}^{1}dx%
\int_{0}^{1}dz\left[ -\frac{(1-z)^{2}}{W}\arcsin {\frac{\omega z}{R}}\right.
\nonumber \\
&&\left. +2\omega ^{2}\cos \theta \frac{x(1-x)z(1-z)^{3}}{W^{3}}\left(
\arcsin \frac{\omega z}{R}+\frac{\omega zW}{R^{2}}\right) \right]
\end{eqnarray}
\begin{eqnarray}
{\cal A}_{6}^{NLO} &=&-\frac{e^{2}\omega }{2M_{N}^{2}}(1+\kappa _{p})+\frac{%
e^{2}g_{A}\omega ^{3}}{8\pi ^{2}F_{\pi }^{2}(m_{\pi }^{2}-t)}+{\frac{%
e^{2}g_{A}^{2}\omega ^{2}}{8\pi ^{2}F_{\pi }^{2}}}\int_{0}^{1}dx%
\int_{0}^{1}dz\left[ \frac{(1-z)^{2}}{W}\arcsin {\frac{\omega z}{R}}\right. 
\nonumber \\
&&\left. -2\omega ^{2}\frac{x(1-x)z(1-z)^{3}}{W^{3}}\left( \arcsin \frac{%
\omega z}{R}+\frac{\omega zW}{R^{2}}\right) \right]
\end{eqnarray}

with 
\begin{gather}
t=\left( k-k^{\prime }\right) ^{2}=-2\omega ^{2}\left( 1-\cos \theta \right)
,  \nonumber \\
W=\sqrt{m_{\pi }^{2}-\omega ^{2}z^{2}+t(1-z)^{2}x(x-1)},\quad R=\sqrt{m_{\pi
}^{2}+t(1-z)^{2}x(x-1)}.
\end{gather}

Expressions of PC Compton scattering for the unpolarized differential cross
section, $d\sigma /d\Omega $, proton-photon spin parallel asymmetry ${\cal A}%
_{\Vert }$, and proton-photon spin perpendicular asymmetry ${\cal A}_{\perp }
$ are given in terms of PC structure functions in eqs.(4.18) and (4.19) of
ref. \cite{BKM}. These structure functions can be extracted experimentally.
For example, below pion production threshold ($\omega <m_{\pi })$, one can
extract ${\cal A}_{1}$ and ${\cal A}_{3}$ by measuring $d\sigma /d\Omega $
and ${\cal A}_{\Vert }$ at $\theta =0$. Since ${\cal A}_{i}$ are all real
below pion production threshold and ${\cal A}_{\Vert }\varpropto {\cal A}%
_{1}^{2}+{\cal A}_{3}^{2}{\cal \ }$and ${\cal A}_{\perp }\varpropto {\cal A}%
_{1}{\cal A}_{3}$, measurements of ${\cal A}_{\Vert }$ and ${\cal A}_{\perp }
$ are sufficient to extract ${\cal A}_{1}$ and ${\cal A}_{3}.$

\bigskip The following PV structure functions are given by ref.\cite{BS}. 
\begin{eqnarray}
{\cal F}_{1}(\omega ,\theta ) &=&{\frac{e^{2}g_{A}h_{\pi NN}^{(1)}}{4\sqrt{2%
}\pi ^{2}F_{\pi }}}\int_{0}^{1}dx\ \int_{0}^{1-x}dy\ (1-2y)\omega \ \left[ 
{\cal I}(-1;x\omega ,\tilde{m}^{2})-{\cal I}(-1;-x\omega ,\tilde{m}^{2})%
\right]   \nonumber \\
{\cal F}_{2}(\omega ,\theta ) &=&{\frac{e^{2}g_{A}h_{\pi NN}^{(1)}}{2\sqrt{2%
}\pi ^{2}F_{\pi }}}\int_{0}^{1}dx\ \int_{0}^{1-x}dy\ y\ \omega \left[ {\cal I%
}(-1;x\omega ,\tilde{m}^{2})-{\cal I}(-1;-x\omega ,\tilde{m}^{2})\right]  
\nonumber \\
{\cal F}_{3}(\omega ,\theta ) &=&{\frac{e^{2}g_{A}h_{\pi NN}^{(1)}}{2\sqrt{2%
}\pi ^{2}F_{\pi }}}\int_{0}^{1}dx\ \int_{0}^{1-x}dy\ y\ (1-x-y)\
(2y-1)\omega ^{3}\ \left[ {\cal I}(-2;x\omega ,\tilde{m}^{2})-{\cal I}%
(-2;-x\omega ,\tilde{m}^{2})\right]   \nonumber \\
\tilde{m}^{2} &=&m_{\pi }^{2}+2y(1-x-y)\ \omega ^{2}\ \left( 1-\cos \theta
\right) \ \ \ ,
\end{eqnarray}
where the functions ${\cal I}(\alpha ;b,c)$ are defined by Jenkins and
Manohar in ref.\cite{HBChPT} 
\begin{eqnarray}
{\cal I}(\alpha ;b,c) &=&\int_{0}^{\infty }d\lambda \ \left( \lambda
^{2}+2\lambda b+c\right) ^{\alpha }  \nonumber \\
{\cal I}(-1;\Delta ,m^{2}) &=&-{\frac{1}{2\sqrt{\Delta ^{2}-m^{2}+i\epsilon }%
}}\log \left( {\frac{\Delta -\sqrt{\Delta ^{2}-m^{2}+i\epsilon }}{\Delta +%
\sqrt{\Delta ^{2}-m^{2}+i\epsilon }}}\right)   \nonumber \\
{\cal I}(-2;\Delta ,m^{2}) &=&{\frac{1}{2\left( \Delta ^{2}-m^{2}+i\epsilon
\right) }}\left( {\frac{\Delta }{m^{2}}}-{\cal I}(-1;\Delta ,m^{2})\right) \
\ \ .
\end{eqnarray}

\end{document}